\documentclass[preprint,authoryear]{elsarticle}
\usepackage{fullpage}
\usepackage{subcaption}
\usepackage{pifont}
\usepackage{latexsym}
\usepackage{mathrsfs}
\usepackage{amssymb,amsbsy}
\usepackage{amsfonts}
\usepackage{graphicx}
\usepackage{tcolorbox}
\usepackage[colorinlistoftodos]{todonotes}
\usepackage{varwidth}
\usepackage{algorithm}
\usepackage{algorithmic}
\usepackage{lmodern}
\usepackage{natbib}
\usepackage{mathtools}          
\usepackage[titletoc]{appendix}
\usepackage{titletoc}
\usepackage{ntheorem}

\def\apj{{\it Astrophys.~J.}}

\def\apjl{{\it Astrophys.~J.~Lett.}}
\def\apss{{\it Astrophys. Space Sci.}}

\def\mnras{{\it Mon.~Not.~R.~Astron.~Soc.} }
\def\nat{{\it Nature} }

\journal{Astronomy and Computing}

\begin{document}

\begin{frontmatter}

\title{CD-HPF: New Habitability Score Via Data Analytic Modeling}

\author[1]{Kakoli Bora}
\author[2]{Snehanshu Saha\corref{cor1}}
\ead{snehanshusaha@pes.edu}
\author[2]{Surbhi Agrawal}
\author[3]{Margarita Safonova}
\author[4]{Swati Routh}
\author[5]{Anand Narasimhamurthy}

\cortext[cor1]{Corresponding author}

\address[1]{Department of Information Science and Engineering, PESIT-BSC, Bangalore}
\address[2]{Department of Computer Science and Engineering, PESIT-BSC, Bangalore}
\address[3]{Indian Institute of Astrophysics, Bangalore}
\address[4]{Department of Physics, Jain University, Bangalore}
\address[5]{BITS, Hyderabad}


\begin{abstract}
The search for life on the planets outside the Solar System can be broadly classified into the following: looking for Earth-like conditions or the planets similar to the Earth (Earth similarity), and looking for the possibility of life in a form known or unknown to us (habitability). The two frequently used indices, Earth Similarity Index (ESI) and Planetary Habitability Index (PHI), describe heuristic methods to score similarity/habitability in the efforts to categorize different exoplanets or exomoons. ESI, in particular, considers Earth as the reference frame for habitability and is a quick screening tool to categorize and measure physical similarity of any planetary body with the Earth. The PHI assesses the probability that life in some form may exist on any given world, and is based on the essential requirements of known life: a stable and protected substrate, energy, appropriate chemistry and a liquid medium. We propose here a different metric, a Cobb-Douglas Habitability Score (CDHS), based on Cobb-Douglas habitability production function (CD-HPF), which computes the habitability score by using measured and calculated planetary input parameters. As an initial set, we used radius, density, escape velocity and surface temperature of a planet. The values of the input parameters are normalized to the Earth Units (EU). The proposed metric, with exponents accounting for metric elasticity, is endowed with verifiable analytical properties that ensure global optima, and is scalable to accommodate finitely many input parameters. The model is elastic, does not suffer from curvature violations and, as we discovered, the standard PHI is a special case of CDHS. Computed CDHS scores are fed to K-NN (K-Nearest Neighbour) classification algorithm  with probabilistic herding that facilitates the assignment of exoplanets to appropriate classes via supervised feature learning methods, producing granular clusters of habitability. The proposed work describes a decision-theoretical model using the power of convex optimization and algorithmic machine learning. 
\end{abstract}

\begin{keyword}
Habitability Score \sep Cobb-Douglas production function \sep expolanets \sep machine learning \sep CDHS \sep optimization
\end{keyword}


\end{frontmatter}

\section{Introduction}

In the last decade, thousands of planets are discovered in our Galaxy alone. The inference is that stars with planets are a rule rather than exception \citep{Cassan2012BoundPlanets}, with estimates of the actual number of planet exceeding the number of stars in our Galaxy by orders of magnitude \citep{Strigari2012Galaxy}. The same line of reasoning suggests a staggering number of at least $10^{24}$ planets in the observable Universe. The biggest question posed therefore is whether there are other life-harbouring planets. The most fundamental interest is in finding the Earth's twin. In fact, {\it Kepler} space telescope  ({\tt{http://kepler.nasa.gov/}}) was designed specifically to look for Earth's analog -- Earth-size planets in the habitable zones (HZ) of G-type stars \citep{Batalha2014Kepler}. More and more evidence accumulated in the last few years suggests that, in astrophysical context, Earth is an average planet, with average chemistry, existing in many other places in the Galaxy, average mass and pressure. Moreover, recent discovery of  the rich organic content in the protoplanetary disk of newly formed star  MWC 480 \citep{O'berg2015CommetLike} has shown that neither is our Solar System unique in the abundance of the key components for life. Yet the only habitable planet in the Universe known to us is our Earth. 

The question of habitability is of such interest and importance that the theoretical work has expanded from just the stellar HZ concept to the Galactic HZ (Gonzales 2001) and, recently, to the Universe HZ --- asking a question which galaxies are more habitable than others (Dayal et al. 2015). However, the simpler  question --- which of thousands detected planets are, or can be, habitable is still not answered. Life on other planets, if exists, may be similar to what we have on our planet, or may be in some other unknown form. The answer to this question may depend on understanding how different physical planetary parameters, such as planet's orbital properties, its chemical composition, mass, radius, density, surface and interior temperature, distance from it's parent star, even parent star's temperature or mass, combine to provide habitable conditions. With currently more than 1800 confirmed and more than 4000 unconfirmed discoveries\footnote{Extrasolar Planets Encyclopedia, http://exoplanet.eu/catalog/}, there is already enormous amount of accumulated data, where the challenge lies in the selection of how much to study about each planet, and which parameters are of the higher priority to evaluate.

Several important characteristics were introduced to address the habitability question. \citet{Schulze2011Two-tired} first addressed this issue through two indices, the Planetary Habitability Index (PHI) and the Earth Similarity Index (ESI), where maximum is set as 1 for a planet where life as we know it had formed; thus for the Earth, PHI$=$ESI$=1$. ESI represents a quantitative measure with which to assess the similarity of a planet with the Earth on the basis of mass, size and temperature. But ESI alone is insufficient to conclude about the habitability, as planets like Mars have ESI close to 0.8 but we cannot still categorize it as habitable. There is also a possibility that a planet with ESI value slightly less than 1 may harbor life in some form which is not there on Earth, i.e. unknown to us. PHI was quantitatively defined as a measure of the ability of a planet to develop and sustain life. However, evaluating PHI values for large number of planets is not an easy task. In \cite{Irwin2014Biological}, another parameter was introduced to account for the chemical composition of exoplanets and some biology-related features such as substrate, energy, geophysics, temperature and age of the planet --- the Biological Complexity Index (BCI). Here, we briefly describe the mathematical forms of these parameters.
 
\paragraph{\bf Earth Similarity Index (ESI)}

ESI was designed to indicate how Earth-like an exoplanet might be \citep{Schulze2011Two-tired} and is an important factor to initially assess the habitability measure. Its value lies between 0 (no similarity) and 1, where 1 is the reference value, i.e. the ESI value of the Earth, and a general rule is that any planetary body with an ESI over 0.8 can be considered an Earth-like. It was proposed in the form 
\begin{equation}
ESI_{x}= \left( 1-\left | \frac{x-x_{0}}{x+x_{0}} \right | \right)^{w}\,,
\label{eq:ESIgeneral}
\end{equation}
where ESI$_x$ is the ESI value of a planet for $x$ property, and $x_{0}$ is the Earth's value for that property. The final ESI value of the planet is obtained by combining the geometric means of individual values, where $w$ is the weighting component through which the sensitivity of scale is adjusted. Four parameters: surface temperature $T_s$, density $D$, escape velocity $V_e$ and radius $R$, are used in ESI calculation. This index is split into interior $ESI_i$ (calculated from radius and density), and surface $ESI_s$ (calculated from escape velocity and surface temperature). Their geometric means are taken to represent the final $ESI$ of a planet. However, ESI in the form (\ref{eq:ESIgeneral}) was not introduced to define habitability, it only describes the similarity to the Earth in regard to some planetary parameters. For example, it is relatively high for the Moon. 

\paragraph{\bf Planetary Habitability Index (PHI)}
 
 To actually address the habitability of a planet, \citet{Schulze2011Two-tired} defined the PHI as
\begin{equation}
PHI=\left(S\cdot E \cdot C\cdot L\right)^{1/4}\,,
\label{eq:PHIgeneral}
\end{equation} where $S$ defines a substrate, $E$ -- the available energy, $C$ -- the appropriate chemistry and $L$ -- the liquid medium; all the variables here are in general vectors, while the corresponding scalars represent the norms of these vectors. For each of these categories, the PHI value is divided by the maximum PHI to provide the normalized PHI in the scale between 0 to 1. However, PHI in the form (\ref{eq:PHIgeneral}) lacks some other properties of a planet which may be necessary for determining its present habitability. For example, in Shchekinov et al. (2013) it was suggested to complement the original PHI with the explicit inclusion of the age of the planet (see their Eq.~6).

\paragraph{\bf Biological Complexity Index (BCI)} 
 
To come even closer to defining habitability, yet another index was introduced, comprising the above mentioned four parameters of the PHI and three extra parameters, such as geophysical complexity $G$, appropriate temperature $T$ and age $A$ \citep{Irwin2014Biological}. Therefore, the total of seven parameters were initially considered to be important for the BCI. However, due to the lack of information on chemical composition and the existence of liquid water on exoplanets, only five were retained in the final formulation,
 \begin{equation}
 BCI=\left(S\cdot E\cdot T\cdot G\cdot A\right)^{1/5}\,.
 \label{eq:BCIgeneral}
 \end{equation}
 It was found in \citet{Irwin2014Biological} that for 5 exoplanets the BCI value is higher than for Mars, and that planets with high BCI values may have low values of ESI. 
 
All previous indicators for habitability assume a planet to reside within in a classical HZ of a star, which is conservatively defined as a region where a planet can support liquid water on the surface \citep{HUANG59,Kasting93}. The concept of an HZ is, however, a constantly evolving one, and it has have been since suggested that a planet may exist beyond the classical HZ and still be a good candidate for habitability \citep{Irwin2011CosmicBiology,Armstrong2014}. Though presently all efforts are in search for the Earth's twin where the ESI is an essential parameter, it never tells that a planet with ESI close to $1$ is habitable. Much advertised recent hype in press about finding the best bet for life-supporting planet -- Gliese 832c with ESI $=0.81$ \citep{Wittenmyer2014GJ832c}, was thwarted by the realization that the planet is more likely to be a super-Venus, with large thick atmosphere, hot surface and probably tidally locked with its star. 
 
We present here the novel approach to determine the habitability score of all confirmed exoplanets analytically. Our goal is to determine the likelihood of an exoplanet to be habitable using the newly defined habitability score (CDHS) based on Cobb-Douglas habitability production function (CD-HPF), which computes the habitability score by using measured and calculated planetary input parameters. Here, the PHI in its original form turned out to be a special case. We are looking for a feasible solution that maximizes habitability scores using CD-HPF with some defined constraints. In the following sections, the proposed model and motivations behind our work are discussed along with the results and applicability of the method. We conclude by listing key takeaways and robustness of the method. The related derivations and proofs are included in the appendices. 
 
\section{CD-HPF: Cobb-Douglas Habitability Production Function}

We first present key definitions and terminologies that are utilized in this paper. These terms play critical roles in understanding the method and the algorithm adopted to accomplish our goal of validating the habitability score, \textbf{CDHS}, by using CD-HPF eventually.

\subsection{Key Definitions}

\begin {itemize}
\item \textbf{Mathematical Optimization}

Optimization is one of the procedures to select the best element from a set of available alternatives in the field of mathematics, computer science, economics, or management science \citep{HAJKOVA2007CD}. An optimization problem can be represented in various ways. Below is the representation of an optimization problem. Given a function \(f:A \rightarrow R\) from a set $A$ to the real numbers $R$. If an element \(x_0\) in $A$ is such that \(f(x_0) \leq f(x)\) for all $x$ in $A$, this ensures minimization. The case \(f(x_0) \geq f(x)\) for all $x$ in $A$ is the specific case of maximization. The optimization technique is particularly useful for modeling the habitability score in our case. In the above formulation, the domain $A$ is called a search space of the function $f$, CD-HPF in our case, and elements of $A$ are called the candidate solutions, or feasible solutions. The function as defined by us is a utility function, yielding the habitability score CDHS. It is a feasible solution that maximizes the objective function, and is called an optimal solution under the constraints known as {\bf Returns to scale}.
\item \textbf{Returns to scale} measure the extent of an additional output obtained when all input factors change proportionally. There are three types of returns to scale: 
\begin{enumerate}
\item{\bf Increasing returns to scale (IRS)}. In this case, the output increases by a larger proportion than the increase in inputs  during the production process. For example, when we multiply the amount of every input by the number $N$, the factor by which output increases is more than $N$. This change occurs as
\begin{enumerate}[(i)]
\item Greater application of the variable factor ensures better utilization of the fixed factor.
\item Better division of the variable factor.
\item It improves coordination between the factors. \end{enumerate}
The 3D plots obtained in this case are neither concave nor convex.
\item{\bf Decreasing returns to scale (DRS)}. Here, the proportion of increase in input increases the output, but in lower ratio, during the production process. For example, when we multiply the amount of every input by the number $N$, the factor by which output increases is less than $N$. This happens because:
\begin{enumerate}[(i)]
\item  As more and more units of a variable factor are combined with the fixed factor, the latter gets over-utilized. Hence, the rate of corresponding growth of output goes on diminishing.
\item Factors of production are imperfect substitutes of each other. The divisibility of their units is not comparable.
\item The coordination between factors get distorted so that marginal product of the variable factor declines.
\end{enumerate}
The 3D plots obtained in this case are concave.
 \item{\bf Constant returns to scale (CRS)}. Here, the proportion of increase in input increases output in the same ratio, during the production process. For example, when we multiply the amount of every input by a number $N$, the resulting output is multiplied by $N$. This phase happens for a negligible period of time and can be considered as a passing phase between IRS and DRS. The 3D plots obtained in this case are concave.
\end{enumerate}
\item \textbf{Computational Techniques in Optimization}. There exist several well-known techniques including Simplex, Newton-like and Interior point-based techniques \citep{optimization}. One such technique is implemented via MATLAB's optimization toolbox using the function
\textbf{fmincon}. This function helps find the global optima of a constrained optimization problem which is relevant to the model proposed and implemented by the authors. Illustration of the function and its syntax are provided in Appendix~D. 
\item \textbf{Concavity}. Concavity ensures global maxima. The implication of this fact in our case is that if CD-HPF is proved to be concave under some constraints (this will be elaborated later in the paper), we are guaranteed to have maximum habitability score for each exoplanet in the global search space.
\item \textbf{Machine Learning}. Classification of patterns based on data is a prominent and critical component of machine learning and will be highlighted in subsequent part of our work where we made use of a standard K-NN algorithm. The algorithm is modified to tailor to the complexity and efficacy of the proposed solution. Optimization, as mentioned above, is the art of finding maximum and minimum of surfaces that arise in models utilized in science and engineering. More often than not, the optimum has to be found in an efficient manner, i.e. both the speed of convergence and the order of accuracy should be appreciably good. Machines are trained to do this job as, most of the times, the learning process is iterative. Machine learning is a set of methods and techniques that are intertwined with optimization techniques. The learning rate could be accelerated as well, making optimization problems deeply relevant and complementary to machine learning. 
\end{itemize}

\subsection{Cobb-Douglas Habitability Production Function CD-HPF}

The general form of the Cobb-Douglas production function CD-PF is 
\begin{equation}          
Y= k \cdot \left(x_{1}\right)^{\alpha}\cdot \left(x_{2}\right)^{\beta} \,,
\label{eq:CDPF}
\end{equation}
where $k$ is a constant that can be set arbitrarily according to the requirement, $Y$ is the total production, i.e. output, which is homogeneous with the degree 1; $x_1$ and $x_2$ are the input parameters (or factors); $\alpha$ and $\beta$ are the real fixed factors, called the elasticity coefficients. The sum of elasticities determines returns to scale conditions in the CDPF. This value can be less than 1, equal to 1, or greater than 1.

What motivates us to use the Cobb-Douglas production function is its properties. Cobb-Douglas production function (Cobb \& Douglas, 1928) was originally introduced for modeling the growth of the American economy during the period of 1899--1922, and is currently widely used in economics and industry to optimize the production while minimizing the costs \citep{De-MinWu1975,Moyazzem2012Cobb,HassaniCobb,Snehanshu2016RevenueOpt}. Cobb-Douglas production function is concave if the sum of the elasticities is not greater than one (see the proof in Bergstrom 2010). This gives global extrema in a closed interval which is handled by constraints in elasticity.
The physical parameters used in the Cobb-Douglas model may change over time and, as such, may be modeled as continuous entities. A functional representation, i.e response, $Y$, is thus a continuous function, and may increase or decrease in maximum or minimum value as these parameters change. Our formulation serves this purpose, where elasticities may be adjusted via {\em fmincon} or fitting algorithms, in conjunction with the intrinsic property of the CD-HPF that ensures global maxima for concavity. Our simulations, that include animation and graphs, support this trend (see Figures 1 and 2 in Section~3). As the physical parameters change in value, so do the function values and its maximum for all the exoplanets in the catalog, and this might rearrange the CDHS pattern with possible changes in the parameters, while maintaining consistency with the database.

The most important properties of this function that make it flexible to be used in various applications are:
\begin{itemize}
\item It can be transformed to the log-linear form from its multiplicative form (non-linear) which makes it simple to handle, and hence, linear regression techniques can be used for  estimation of missing data.
\item Any proportional change in any input parameter can be represented easily as the change in the output.
\item The ratio of relative inputs $x_1$ and $x_2$ to the total output $Y$ is represented by the elasticities $\alpha$ and $\beta$.
\end{itemize}
The analytical properties of the CD-HPF motivated us to check the applicability in our problem, where the four parameters considered to estimate the habitability score are surface temperature, escape velocity, radius and density. Here, the production function $Y$ is the habitability score of the exoplanet, where the aim is to maximize $Y$, subject to the constraint that the sum of all elasticity coefficients shall be less than or equal to 1. Computational optimization is relevant for elasticity computation in our problem. Elasticity is the percentage change in the output $Y$, given one percent change in the input parameter, $x_1$ or $x_2$. We assume $k$ is constant. In other words, we compute the rate of change of output $Y$, the CD-HPF, with respect to one unit of change in input, such as $ x_1$ or $x_2$. As the quantity of $x_1$ or $ x_2$ increases by one percent, output increases by $\alpha$ or $\beta$ percent. This is known as the elasticity of output with respect to an input parameter. As it is, values of the elasticity, $\alpha$ and $\beta$ are not ad-hoc and need to be approximated for optimization purpose by some computational technique. The method, {\em fmincon} with interior point search, is used to compute the elasticity values for CRS, DRS and IRS. The outcome is quick and accurate. We elaborate the significance of the scales and elasticity in the context of CD-HPF and CDHS below.  
\\
\begin{itemize} 
\item \textbf{Increasing returns to scale (IRS):} In Cobb-Douglas model, if $\alpha+\beta>1$, the case is called an IRS. It improves the coordination among the factors. This is indicative of boosting the habitability score following the model with one unit of change in respective predictor variables. 
\item \textbf{Decreasing returns to scale (DRS):} In Cobb-Douglas model, if $\alpha+\beta<1$, the case is called a DRS, where the deployment of an additional input may affect the output with diminishing rate. This implies the habitability score following the model may decrease with the one unit of change in respective predictor variables.
\item \textbf{Constant returns to scale (CRS):} In Cobb-Douglas model, if $\alpha+\beta=1$, this case is called a CRS, where increase in $\alpha$ or/and $\beta$ increases the output in the same proportion. The habitability score, i.e the response variable in the Cobb-Douglas model, grows proportionately with changes in input or predictor variables.
\end{itemize}
The range of elasticity constants is between 0 and 1 for DRS and CRS. This will be exploited during the simulation phase (Section~3). It is proved in Appendices B and C that the habitability score  (CDHS) maximization is accomplished in this phase for \textbf{DRS and CRS},  respectively.

The impact of change in the habitability score according to each of the above constraints will be elaborated in Sections $4$ and $5$. Our aim is to optimize elasticity coefficients to maximize  the habitability score of the confirmed exoplanets using the CD-HPF.

\subsection{Cobb-Douglas Habitability Score estimation}

We have considered the same four parameters used in the ESI metric (Eq.~\ref{eq:ESIgeneral}), i.e. surface temperature, escape velocity, radius and density, to calculate the Cobb-Douglas Habitability Score (CDHS). Analogous to the method used in ESI, two types of Cobb-Douglas Habitability Scores are calculated -- the interior CDHS$_i$ and the surface CDHS$_i$. The final score is computed by a linear convex combination of these two, since it is well known that a convex combination of convex/concave function is also convex/concave. The interior CDHS$_i$, denoted by $Y1$, is calculated using radius $R$ and density $D$, 
\begin{equation} 
Y1=CDHS_{i}=(D)^{\alpha}\cdot (R)^{\beta} \,.
\end{equation}
The surface CDHS$_s$, denoted by $Y2$, is calculated using surface temperature $T_s$ and escape velocity $V_e$, 
\begin{equation}
Y2=CDHS_{s}=\left(T_{s}\right)^{\gamma}\cdot \left(V_{e}\right)^{\delta}\,. 
\end{equation}
The final CDHS $Y$, which is a convex combination of $Y1$ and $Y2$, is determined by
\begin{equation}
Y=w^{\prime} \cdot Y1 + w^{\prime\prime} \cdot Y2\,,
\end{equation}
where the sum of $w^{\prime}$ and $w^{\prime\prime}$ equals $1$. The values of $w^{\prime}$ and $w^{\prime\prime}$ are the weights of the interior CDHS$_i$ and surface CDHS$_s$, respectively. These weights depend on the importance of individual parameters of each exoplanet. The $Y1$ and $Y2$ are obtained by applying CD-HPF (Eq.~\ref{eq:CDPF})
with $k=1$. Finally, the Cobb-Douglas habitability production function can be formally written as 
\begin{equation}
Y=f\left(R,D,T_{s},V_{e}\right)=\left(R\right)^{\alpha}\cdot \left(D\right)^{\beta}
\cdot \left(T_{s}\right)^{\gamma}\cdot\left(V_{e}\right)^{\delta}\,.
\end{equation}  
The goal is to maximize $Y$, iff $\alpha+\beta+\gamma+\delta<1$. The ease of visualization is the reason \textbf{CDHS} is computed by splitting into two parts $Y1$ and $Y2$ and combining by using the weights $w^{\prime}$ and $w^{\prime\prime}$. Individually, each of $Y1$ and $Y2$ are sample 3D models and, as such, are easily comprehensible via surface plots as demonstrated later (see Figs.~1 and 2 in Section~3). The authors would like to emphasize that instead of splitting and computing $Y$ as a convex combination of $Y1$ and $Y2$, a direct calculation of $Y$ is possible, which does not alter the final outcome. It is avoided here, since using the product of all four parameters with corresponding elasticities $\alpha,\beta,\gamma$ and $\delta$ would make rendering the plots impossible for the simple reason of dimensionality being too high, $5$ instead of $3$. We reiterate that the scalability of the model from $\alpha,\beta$ to $\alpha,\beta,\gamma$ and $\delta$ does not suffer due to this scheme. The proof presented in Appendix~B bears testimony to our claim.

\subsection{The Theorem for Maximization of Cobb-Douglas habitability production function}
\textbf{Statement:} CD-HPF attains global maxima in the phase of DRS or CRS \citep{Snehanshu2016RevenueOpt}.\\
{\em Sketch of proof}: Generally profit of a firm can be defined as
\begin{equation}		
\mbox{profit} = \mbox{revenue} - \mbox{cost} = \left(\mbox{price of output} \times \mbox{output}\right) - \left(\mbox{price  of input}\times \mbox{input}\right)\,.\nonumber
\end{equation}

Let $p_{1},p_{2},\ldots,p_{n}$ be a vector of prices for outputs, or products, and $w_{1},w_{2},\ldots,w_{m}$ be a vector of prices for inputs of the firm, which are always constants; and let the input levels be $x_{1},x_{2},\ldots,x_{m}$, and the output levels be $y_{1},y_{2},\ldots,y_{n}$. The profit, generated by the production plan, ($x_1,\dots,x_m,y_1,\dots,y_n$) is  
\begin {equation}
\pi =\left(p_{1}\cdot y_{1}+ \ldots +p_{n}\cdot y_{n} - w_{1}\cdot x_{1} - \ldots - w_{m}\cdot x_{m}\right)\,.
\nonumber 
\end{equation}
Suppose the production function for $m$ inputs is 
\begin{equation}
Y = f\left(x_{1},x_{2},...,x_{m}\right)\,,
\nonumber 
\end{equation}
and its profit function is
\begin{equation}
\pi = p\cdot Y - w_{1}\cdot x_{1}- \ldots - w_{m}\cdot x_{m}\,.
\nonumber 
\end{equation}
A single output function needs $p$ as the price, while multiple output functions will require multiple prices $p_1, p_2, \ldots, p_n$. The profit function in our case, which is a single-output multiple-inputs case, is given by
\begin{equation}
\pi=pf\left(R,D,T_{s},V_{e}\right)-w_1 R-w_2 D-w_3 T_s - w_4 V_e \,,
\end{equation}
where $w_{1}\),\(w_{2}\),\(w_{3}\),\(w_{4}$ are the weights chosen according to the importance for habitability for each planet. Maximization of CD-HPF is achieved when 
\begin{equation}
(1)\,\, p\frac{\partial f}{\partial R}=w_1\,,\quad  (2)\,\,  p\frac{\partial f}{\partial D}=w_2\,,\quad  (3) \,\, p\frac{\partial f}{\partial T_s}=w_3 \,,\quad (4)\, \, p\frac{\partial f}{\partial V_e}=w_4\,.
\end{equation}
The habitability score is conceptualized as a profit function where the cost component is introduced as a penalty function to check unbridled growth of CD-HPF. This bounding framework is elaborated in the proofs of concavity, the global maxima and computational optimization technique, and function {\em fmincon} in Appendices B, C and D, respectively. \\

{\bf Remark}: If we consider the case of CRS, where all the elasticities of different cost components are equal, the output is $Y=\prod_{i=1}^{n} x_i^{\alpha_i}$, where all $\alpha_i$ are equal and $\sum\alpha_i=1$. In such scenario, $Y\equiv G.M.$ (Geometric Mean) of the cost inputs. Further scrutiny reveals that the geometric mean formalization is nothing but the representation of the PHI, thus establishing our framework of CD-HPF as a broader model, with the PHI being a corollary for the CRS case.

Once we compute the habitability score, $Y$, the next step is to perform clustering of the Y values. We have used K-nearest neighbor (K-NN) classification algorithm and introduced probabilistic herding and thresholding to group the exoplanets according to their $Y$ values. The algorithm finds the exoplanets for which $Y$ values are very close to each other and keeps them in the same group, or cluster. Each CDHS value is compared with its $K$ (specified by the user) nearest exoplanet's (closer $Y$ values) CDHS value, and the class which contains maximum nearest to the new one is allotted as a class for it.

\section{Implementation of the Model}

We applied the CD-HPF to calculate the Cobb-Douglas habitability score (CDHS) of exoplanets. A total of 664 confirmed exoplanets are taken from the Planetary Habitability Laboratory Exoplanets Catalog (PHL-EC)\footnote{provided by the Planetary Habitability Laboratory @ UPR Arecibo, accessible at \\{\tt http://phl.upr.edu/projects/habitable-exoplanets-catalog/data/database}}. The catalog contains observed and estimated stellar and planetary parameters for all currently confirmed exoplanets. We have used only those entries, for which the mean surface temperature was provided --- 664 planets (only planets with solid surfaces, according to PHL-EC, including Earth). As mentioned above, the CDHS of exoplanets are computed from the interior CDHS$_i$ and the surface CDHS$_s$. The input parameters radius $R$ and density $D$ are used to compute the values of the elasticities \(\alpha\) and \(\beta\). Similarly, the input parameters surface temperature $T_S$ and escape velocity $V_e$ are used to compute the elasticities \(\gamma\) and \(\delta\). These parameters, except the surface temperature, are given in Earth Units (EU) in the PHL-EC catalog. We have normalized the surface temperatures $T_s$ of exoplanets to the EU, by dividing each of them with Earth's mean surface temperature, 288 K. 

The Cobb-Douglas function is applied on varying elasticities to find the CDHS close to Earth's value, which is considered as $1$. As all the input parameters are represented in EU, we are looking for the exoplanets whose CDHS is closer to Earth's CDHS. For each exoplanet, we obtain the optimal elasticity and the maximum CDHS value. The results are demonstrated graphically using 3D plot. All simulations were conducted using the MATLAB software for the cases of DRS and CRS. From Eq.~(B.38), we can see that for CRS $Y$ will grow asymptotically, if 
\begin{equation} 
\alpha+ \beta + \gamma + \delta = 1\,. \end{equation}
Let us set 
\begin{equation} 
\alpha=\beta=\gamma=\delta=1/4\,. \end{equation}
In general, the values of elasticities may not be equal but the sum may still be 1. As we know already, this is CRS. A special case of CRS, where the elasticity values are made to be equal to each other in Eq.~(12), turns out to be structurally analogous to the PHI and BCI formulations. Simply stated, the CD-HPF function satisfying this special condition may be written as
\begin{equation}
Y=f=k\left(R\cdot D\cdot T_{s}\cdot V_{e}\right)^{1/4}\,. 
\end{equation}
The function is concave for CRS and DRS (Appendices~B and C). 

\subsection{Computation of CDHS in DRS phase} 

We have computed elasticities separately for interior CDHS$_i$ and surface CDHS$_s$ in the DRS phase. These values were obtained using function {\em fmincon}, a computational optimization technique explained in Appendix D. Tables~1 through~3 show a sample of computed values. Table~\ref{table:interior} shows the computed elasticities $\alpha,\beta$ and CDHS$_i$. The optimal interior CDHS$_i$ for most exoplanets are obtained at $\alpha = 0.8$ and $\beta = 0.1$. Table~2 shows the computed elasticities $\gamma, \delta$ and CDHS$_{s}$. The optimal surface CDHS are obtained at $\gamma = 0.8$ and $\delta = 0.1$. Using these results, 3D graphs are generated and are shown in Figure~1. The $X$ and $Y$ axes represent elasticities and $Z$-axis represents CDHS of exoplanets. The final CDHS, $Y$, calculated using Eq.~(7) with $w^{\prime}=0.99$ and $w^{\prime\prime}= 0.01$, is presented in Table~3.  
\begin{table}[hb!]
\caption{Sample simulation output of interior CDHS$_i$ of exoplanets calculated from radius and density for DRS}
\centering
\begin{tabular}{c c c c c c} 
\hline 
Exoplanet&Radius&Density&Elasticity($\alpha$)&Elasticity ($\beta$)& CDHS$_{i}$ \\
[0.5ex]
\hline
GJ 163 c & 1.83 & 1.19 & 0.8 & 0.1 & 1.65012\\
GJ 176 b & 1.9  & 1.23 & 0.8 & 0.1 & 1.706056\\
GJ 667C b &	1.71 & 1.12 & 0.8 & 0.1 & 1.553527\\
GJ 667C c	& 1.54	& 1.05 & 0.8 & 0.1 & 1.4195\\
GJ 667C d & 1.67 & 1.1 & 0.8 & 0.1 & 1.521642\\
GJ 667C e & 1.4 & 0.99 & 0.8 & 0.1 & 1.307573\\
GJ 667C f & 1.4 & 0.99 & 0.8 & 0.1 & 1.307573\\
GJ 3634 b & 1.81 & 1.18 & 0.8 & 0.1 & 1.634297\\
Kepler-186 f & 1.11 & 0.9 & 0.8 & 0.1 & 1.075679\\
Gl 15 A b & 1.69 & 1.11 & 0.8 & 0.1 & 1.537594\\
HD 20794 c & 1.35 & 0.98 & 0.8 & 0.1 & 1.26879\\
HD 40307 e & 1.5 & 1.03 & 0.8 & 0.1 & 1.387256\\
HD 40307 f & 1.68 & 1.11 & 0.8 & 0.1 & 1.530311\\
HD 40307 g & 1.82 & 1.18 & 0.8 & 0.1 & 1.641517\\ \hline
\end{tabular}
\label{table:interior}
\end{table}

\begin{table}[ht!]
\caption{Sample simulation output of surface CDHS of exoplanets calculated from  escape velocity and surface temperature for DRS}
\centering
\begin{tabular}{c c c c c c} 
\hline 
 Exoplanet & Escape Velocity & Surface temperature & Elasticity ($\gamma$) & Elasticity ($\delta$) & CDHS$_{s}$ \\
[0.5ex]
\hline
GJ 163 c & 1.99 & 1.11146 & 0.8 & 0.1 & 1.752555\\
GJ 176 b & 2.11 & 1.67986 & 0.8 & 0.1 & 1.91405\\
GJ 667C b & 1.81 & 1.49063 & 0.8 & 0.1 & 1.672937\\
GJ 667C c & 1.57 & 0.994 & 0.8 & 0.1 & 1.433764\\
GJ 667C d & 1.75 & 0.71979 & 0.8 & 0.1 & 1.51409\\
GJ 667C e & 1.39 & 0.78854 & 0.8 & 0.1 & 1.27085\\
GJ 667C f & 1.39 & 0.898958 & 0.8 & 0.1 & 1.287614\\
GJ 3634 b & 1.97 & 2.1125 & 0.8 & 0.1 & 1.946633\\
Kepler-186 f & 1.05 & 0.7871 & 0.8 & 0.1 & 1.015213\\
 Gl 15 A b & 1.78 & 1.412153 & 0.8 & 0.1 & 1.641815\\
HD 40307 e & 1.53 & 1.550694 & 0.8 & 0.1 & 1.482143\\
HD 40307 f & 1.76 & 1.38125 & 0.8 & 0.1 & 1.623444\\
HD 40307 g & 1.98 & 0.939236 & 0.8 & 0.1 & 1.716365\\
HD 20794 c & 1.34 & 1.89791667 & 0.8 & 0.1 & 1.719223\\
\hline
\end{tabular}\label{table:surface}
\end{table}\nopagebreak

\begin{figure}[h!]
\includegraphics[width=0.55\textwidth,height=0.35\textheight]{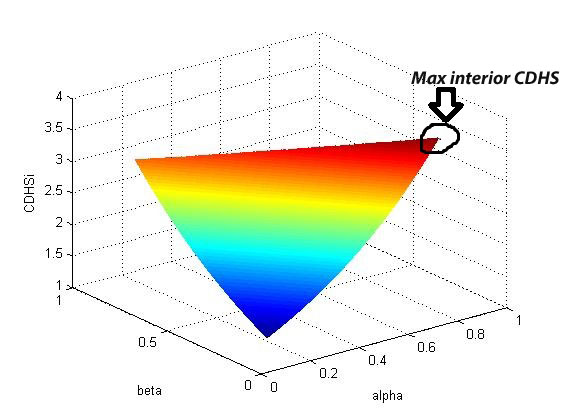}
\includegraphics[width=0.5\textwidth,height=0.35\textheight]{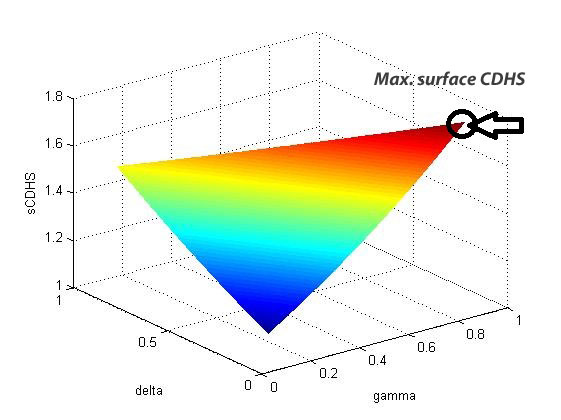}
\caption{Plot of interior CDHS$_i$ ({\it Left}) and surface CDHS$_s$ ({\it Right}) for DRS}

\end{figure}

\begin{table}[ht!]
\caption{Sample simulation output of CDHS with $w^{\prime}=0.99$ and $w^{\prime\prime}=0.01$ for DRS}
\centering
\begin{tabular}{c c c c} 
\hline 
 Exoplanet & CDHS$_i$ & CDHS$_s$ & CDHS \\
[0.5ex]
\hline
GJ 163 c & 1.65012 & 1.752555 & 1.651144\\
GJ 176 b & 1.706056 & 1.91405	& 1.708136\\
GJ 667C b & 1.553527 & 1.672937 & 1.554721\\
GJ 667C c & 1.4195 & 1.433764 & 1.419643\\
GJ 667C d & 1.521642 & 1.514088 & 1.521566\\
GJ 667C e & 1.307573 & 1.27085	& 1.307206\\
GJ 667C f & 1.307573 & 1.287614	& 1.307373\\
GJ 3634 b & 1.634297 & 1.946633	 & 1.63742\\
Gl 15 A b & 1.537594 & 1.641815 & 1.538636\\
Kepler-186 f & 1.075679 & 1.015213 & 1.075074\\
HD 20794 c & 1.26879 & 1.719223	& 1.273294\\
HD 40307 e & 1.387256 &	1.482143 &	1.388205\\
HD 40307 f & 1.530311 & 1.623444 & 1.531242\\
HD 40307 g & 1.641517 7 & 1.716365 & 1.642265\\
\hline
\end{tabular}
\end{table}

\subsection{Computation of CDHS in CRS phase}

The same calculations were carried out for the CRS phase. Tables~4, 5 and 6 show the sample of computed elasticities and  habitability scores in CRS. The convex combination of CDHS$_i$ and CDHS$_s$ gives the final CDHS (Eq.~7) with $w^{\prime}=0.99$ and $w^{\prime\prime}=0.01$. The optimal interior CDHS$_i$ for most exoplanets were obtained at $\alpha = 0.9$ and $\beta = 0.1$, and the optimal surface CDHS$_s$ were obtained at $\gamma = 0.9$ and $\delta = 0.1$. Using these results, 3D graphs were generated and are shown in Figure~2. 

\begin{table}[h!]
\caption{Sample simulation output of interior CDHS$_i$ of exoplanets calculated from radius and density for CRS}
\centering
\begin{tabular}{c c c c c c} 
\hline 
 Exoplanet&Radius&Density&Elasticity\((\alpha\))&Elasticity\ \((\beta\))&\(CDHS_{i}\) \\
[0.5ex]
\hline
GJ 163 c & 1.83 & 1.19 & 0.9 & 0.1 & 1.752914\\
GJ 176 b & 1.9  & 1.23 & 0.9 & 0.1 & 1.819151\\
GJ 667C b &	1.71 & 1.12 & 0.9 & 0.1 & 1.639149\\
GJ 667C c	& 1.54	& 1.05 & 0.9 & 0.1 & 1.482134\\
GJ 667C d & 1.67 & 1.1 & 0.9 & 0.1 & 1.601711\\
GJ 667C e & 1.4 & 0.99 & 0.9 & 0.1 & 1.352318\\
GJ 667C f & 1.4 & 0.99 & 0.9 & 0.1 & 1.352318\\
GJ 3634 b & 1.81 & 1.18 & 0.9 & 0.1 & 1.734199\\
Kepler-186 f & 1.11 & 0.9 & 0.9 & 0.1 & 1.086963\\
Gl 15 A b & 1.69 & 1.11 & 0.9 & 0.1 & 1.62043\\
HD 20794 c & 1.35 & 0.98 & 0.9 & 0.1 & 1.307444\\
HD 40307 e & 1.5 & 1.03 & 0.9 & 0.1 & 1.444661\\
HD 40307 f & 1.68 & 1.11 & 0.9 & 0.1 & 1.611798\\
HD 40307 g & 1.82 & 1.18 & 0.9 & 0.1 & 1.74282\\ \hline
\end{tabular}
\end{table}

\begin{table}[ht!]
\caption{Sample simulation output of surface CDHS of exoplanets calculated from escape velocity and surface temperature for CRS}
\centering
\begin{tabular}{c c c c c c} 
\hline 
 Exoplanet & Escape Velocity & Surface temperature & Elasticity ($\gamma$) & Elasticity ($\delta$) & CDHS$_s$ \\
[0.5ex]
\hline
GJ 163 c & 1.99 & 1.11146 & 0.9 & 0.1 & 1.877401\\
GJ 176 b & 2.11 & 1.67986 & 0.9 & 0.1 & 2.062441\\
GJ 667C b & 1.81 & 1.49063 & 0.9 & 0.1 & 1.775201\\
GJ 667C c & 1.57 & 0.994 & 0.9 & 0.1 & 1.499919\\
GJ 667C d & 1.75 & 0.71979 & 0.9 & 0.1 & 1.601234\\
GJ 667C e & 1.39 & 0.78854 & 0.9 & 0.1 & 1.313396\\
GJ 667C f & 1.39 & 0.898958 & 0.9 & 0.1 & 1.330722\\
GJ 3634 b & 1.97 & 2.1125 & 0.9 & 0.1 & 2.097798\\
Kepler-186 f & 1.05 & 0.7871 & 0.9 & 0.1 & 1.020179\\
 Gl 15 A b & 1.78 & 1.412153 & 0.9 & 0.1 & 1.739267\\
HD 40307 e & 1.53 & 1.550694 & 0.9 & 0.1 & 1.548612\\
HD 40307 f & 1.76 & 1.38125 & 0.9 & 0.1 & 1.717863\\
HD 40307 g & 1.98 & 0 .939236 & 0.9 & 0.1 & 1.837706\\
HD 20794 c & 1.34 & 1.89791667 & 0.9 & 0.1 & 1.832989\\
\hline
\end{tabular}
\end{table}

\begin{figure}[hb!]
\includegraphics[width=0.53\textwidth,height=0.35\textheight]{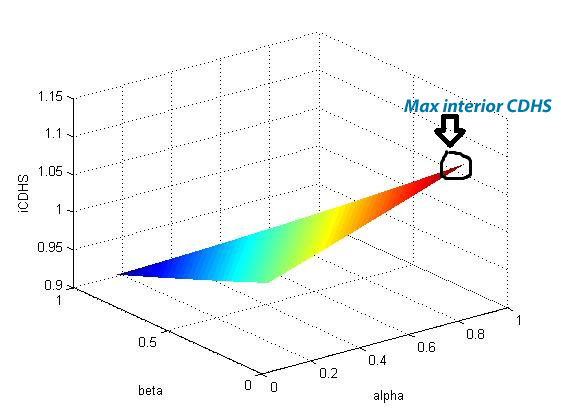}
\includegraphics[width=0.55\textwidth,height=0.35\textheight]{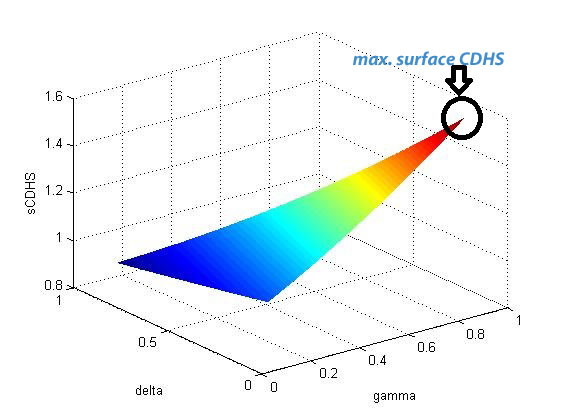}
\caption{Plot of interior CDHS$_i$ ({\it Left}) and surface CDHS$_s$ ({\it Right}) for CRS}
\end{figure}

\begin{table}[h!]
\caption{Sample simulation output of CDHS with $w^{\prime}=0.99$ and $w^{\prime\prime}=0.01$ for CRS}
\centering
\begin{tabular}{c c c c} 
\hline 
 Exoplanet & CDHS$_i$ & CDHS$_s$ & CDHS \\
[0.5ex]
\hline
GJ 163 c & 1.752914	& 1.877401	& 1.754159\\
GJ 176 b & 1.819151	& 2.062441	& 1.821584\\
GJ 667C b & 1.639149 & 1.775201	& 1.64051\\
GJ 667C c & 1.482134	& 1.499919	& 1.482312\\
GJ 667C d & 1.601711	& 1.601234	& 1.601706\\
GJ 667C e & 1.352318 & 1.313396	& 1.351929\\
GJ 667C f & 1.352318 & 1.330722	& 1.352102\\
GJ 3634 b & 1.734199 & 2.097798	& 1.737835\\
Kepler-186 f & 1.086963	& 1.020179	& 1.086295\\
GI 15 A b & 1.62043	& 1.739267	& 1.621618\\
HD 40307 e & 1.444661	& 1.548612 & 1.445701\\
HD 40307 f & 1.611798 & 1.717863 & 1.612859\\
HD 40307 g & 1.74282 & 1.837706	& 1.743769\\
HD 20794 c & 1.307444	& 1.832989	& 1.312699\\
\hline
\end{tabular}
\end{table}

\vskip 0.1in

Tables~1,~2 and 3 represent CDHS for DRS, where the corresponding values of elasticities were found by {\em fmincon} to be 0.8 and 0.1, and the sum$ =0.9 <1$ (The theoretical proof is given in Appendix~B).  Tables~4, 5 and 6 show results for CRS, where the sum of the elasticities $ = 1$ (The theoretical proof is given in Appendix~C). The approximation algorithm {\em fmincon} initiates the search for the optima by starting from a random initial guess, and then it applies a step increment or decrements based on the gradient of the function based on which our modeling is done. It terminates when it cannot find elasticities any better for the maximum CDHS. The plots in Figures~1 and 2 show all the elasticities for which {\em fmincon} searches for the global maximum in CDHS, indicated by a black circle. Those values are read off from the code (given in Appendix~E) and printed as 0.8 and 0.1, or whichever the case may be. A minimalist web page is designed to host all relevant data and results: sets, figures, animation video and a graphical abstract. It is available at {\tt https://habitabilitypes.wordpress.com/}. 
 
The animation video, available at the website, demonstrates the concavity property of CD-HPF and CDHS. The animation comprises 664 frames (each frame is a surface plot essentially), corresponding to 664 exoplanets under consideration. Each frame is a visual representation of the outcome of CD-HPF and CDHS applied to each exoplanet. The $X$ and $Y$ axes of the 3D plots represent elasticity constants and $Z$-axis represents the CDHS. Simply stated, each frame, demonstrated as snapshots of the animation in Figures~1 and 2, is endowed with a maximum CDHS and the cumulative effect of all such frames is elegantly captured in the animation.

\subsection{Attribute Enhanced K-NN Algorithm: A Machine learning approach}

K-NN, or K-nearest neighbor, is a well-known machine learning algorithm. Attribute enhanced K-NN algorithm is used to classify the exoplanets into different classes based on the computed CDHS values. The algorithm produces 6 classes, wherein each class carries exoplanets with CDHS values close to each other, a first condition for being called as "neighbours". 80\% of data from the Habitable Exoplanets Catalog (HEC)\footnote{The Habitable Exoplanets Catalog (HEC) is an online database of potentially habitable planets, total 32 as on January 16, 2016; maintained by the Planetary Habitability Laboratory\@UPR Arecibo, and available at {\tt http://phl.upr.edu/projects/habitable-exoplanets-catalog}}) are used for training, and remaining 20\% for testing. Training--testing process is integral to machine learning, where the machine is trained to recognize patterns by assimilating a lot of data and, upon applying the learned patterns, identifies new data with a reasonable degree of accuracy. The efficacy of a learning algorithm is reflected in the accuracy with which the test data is identified. The training data set is uniformly distributed between first 5 classes, known as balancing the data, so that bias in the training sample is eliminated. Initially, each class holds one fifth of the training data and a new class, i.e. Class 6, defined as Earth's Class (or "Earth-League"), has been derived by the proposed algorithm from first 5 classes where it contains data based on the two conditions. \\ 
\newline  
The two conditions that our algorithm uses to select exoplanets into Class 6 are defined as:
\begin{enumerate}
\item{Thresholding}: Exoplanets with their CDHS minus Earth's CDHS being less than or equal to the specified boundary value, called threshold. We have set a threshold in such a way that the exoplanets with CDHS values within the threshold of 1 (closer to Earth) fall in Earth's class. The threshold is chosen to capture proximal planets as the CDHS of all exoplanets considered vary greatly 

However, this proximity alone does not determine habitability. 
\item{Probabilistic Herding}: if exoplanet is in the HZ of its star, it implies probability of membership to the Earth-League, Class 6, to be high; probability is low otherwise. Elements in each class in K-NN get re-assigned during the run time. This automatic re-assignment of exoplanets to different classes is based on a weighted likelihood concept applied on the members of the initial class assignment.
\end{enumerate}
Consider $K$ as the desired number of nearest neighbors and let $S:=p_{1},\dots,p_{n}$ be the set of training samples in the form $p_{i}=(x_{i},c_{i})$, where \(x_{i}\) is the $d$-dimensional feature vector of the point \(p_{i}\) and \(c_{i}\) is the class that \(p_{i}\) belongs to. In our case, dimension, $d=1$. We fix $S':=p_{1'},\dots,p_{m'}$ to be the set of testing samples. For every sample, the difference in CDHS between Earth and the sample is computed by looping through the entire dataset containing the $5$ classes. Class 6 is the offspring of these 5 classes and is created by the algorithmic logic to store the selected exoplanets which satisfy the conditions of the K-NN and the two conditions -- thresholding and probabilistic herding defined above. We train the system for 80\% of the data-points based on the two constraints, prob$(habitability_{i}) =$ `high' and  CDHS($p_{i}$)$-$CDHS(\textbf{Earth})$\leq$ threshold. These attributes enhance the standard K-NN and help the re-organization of 
\noindent\(exoplanet_{i}\) to Class 6.  \\

If CDHS of \(exoplanet_{i}\) falls with a certain range, \noindent K-NN classifies it accordingly into one of the remaining 5 classes. For each $p'=(x',c')$, we compute the distance $d(x',x_{i})$ between $p'$ and all $p_{i}$ for the dataset of 664 exoplanets from the PHL-EC, $S$. Next, the algorithm selects the $K$ nearest points to $p'$ from the list computed above. The classification algorithm, K-NN, assigns a class $c'$ to $p'$ based on the condition  prob($habitability_{i}) =$ `high' plus the thresholding condition mentioned above. Otherwise, K-NN assigns $p'$ to the class according to the range set for each class. Once the "Earth-League" class is created after the algorithm has finished its run, the list is cross-validated with the habitable exoplanet catalog HEC. It must be noted that Class 6 not only contains exoplanets that are similar to Earth, but also the ones which are most likely to be habitable. The algorithmic representation of K-NN is presented in Appendix E.

\section{\textbf{Results and Discussion}}

The K-NN classification method has resulted in "Earth-league", Class 6, having 14 and 12 potentially habitable exoplanets by DRS and CRS computations, respectively. The outcome of the classification algorithm is shown in Tables~7 and 8.

\begin{table}[h!]
\parbox{.45\linewidth}{
\begin{center}
\caption{Potentially habitable exoplanets in Earth's class using DRS: Outcome of CDHS and K-NN}
\begin{tabular}{cc} 
\hline
Exoplanet&CDH Score\\
[0.5ex]
\hline
GJ 667C e &	1.307206\\
GJ 667C f & 1.307373\\
GJ 832 c & 1.539553 \\
HD 40307 g & 1.642265\\
Kapteyn's b & 1.498503\\
Kepler-61 b & 1.908765\\
Kepler-62 e	& 1.475502 \\
Kepler-62 f	& 1.316121 \\
Kepler-174 d & 1.933823 \\
Kepler-186 f & 1.07507\\
Kepler-283 c & 1.63517\\
Kepler-296 f & 1.619423\\
GJ 667C c & 1.419643 \\
GJ 163 c & 1.651144 \\
\hline
\end{tabular}
\end{center}}
\hfill
\parbox{.45\linewidth}{
\begin{center}
\caption{Potentially habitable exoplanets in Earth's class using CRS: Outcome of CDHS and K-NN}
\begin{tabular}{cc} 
\hline 
 Exoplanet&CDH Score \\
[0.5ex]
\hline
GJ 667C e &	1.351929 \\
GJ 667C f & 1.352102 \\
GJ 832 c & 1.622592 \\
HD 40307 g & 1.743769\\
Kapteyn's b & 1.574564\\
Kepler-62 e	& 1.547538\\
Kepler-62 f	& 1.362128 \\
Kepler-186 f & 1.086295	\\
Kepler-283 c & 1.735285\\
Kepler-296 f & 1.716655	\\
GJ 667C c & 1.482312 \\
GJ 163 c & 1.754159	\\
\hline
\end{tabular}
\end{center}}
\end{table} 

There are 12 common exoplanets in Tables~7 and 8. 
We have cross-checked these planets with the Habitable Exoplanets Catalog and found that they are indeed listed as potentially habitable planets. Class 6 includes all the exoplanets whose CDHS is proximal to Earth. As explained above, classes 1 to 6 are generated by the machine learning technique and classification method. Class 5 includes the exoplanets which are likely to be habitable, and planets in Classes 1,~2,~3 \& 4 are less likely to be habitable, with Class 1 being the least likely to be habitable. Accuracy achieved here is 92\% for $K=1$, implying 1-nearest neighbor, and is 94\% for $K=7$, indicating 7 nearest neighbors. 

In Figure~3 we show the plots of K-NN algorithm applied on the results in DRS (top plot) and CRS (bottom plot) cases. The $X$-axis represents CDHS and $Y$-axis -- the 6 different classes assigned to each exoplanet. The figure is a schematic representation of the outcome of our algorithm. The color points, shown in circles and boxes to indicate the membership in respective classes, are representative of membership only and do not indicate a quantitative equivalence. The numerical data on the number of the exoplanets in each class is provided in Appendix~F. A quantitative representation of the figures may be found at {https://habitabilitypes.wordpress.com/}.  

We also normalized CDHS of each exoplanet, dividing by the maximum score in each category, for both CRS and DRS cases. This resulted in CDHS of all $664$ exoplanets ranging from 0 to 1. Analogous to the case of non-normalized CDHS, these exoplanets have been assigned equally to $5$ classes. K-NN algorithm was then applied to all the exoplanets' CDHS for both CRS and DRS cases. Similar to the method followed in non-normalized CDHS for CRS and DRS, K-NN has been applied to "dump" exoplanets which satisfy the criteria of being members of Class 6. Table~9 shows the potentially habitable exoplanets obtained from classification on normalized data for both CRS and DRS. This result is illustrated in Figs.~3c and 3d. In this figure, Class 6 contains 16 exoplanets generated by K-NN and which are considered to be potentially habitable according to the PHL-EC. The description of the remaining classes is the same as in Figs.~3a and 3b. 

\begin{table}[!ht ]
\centering
\caption{The outcome of K-NN on normalized dataset: potentially habitable exoplanets in Class 6 (Earth-League).}
\begin{tabular}{c c c} 
\hline 
Exoplanet&DRSnormCDHS&CRSnormCDHS \\
[0.5ex]
\hline
GJ 667C e & 0.007833698 &	0.004294092\\
GJ 667C f & 0.007834698 & 0.004294642\\
GJ 832 c & 0.009226084  & 0.005153791 \\
HD 40307 g & 0.009841607 & 0.005538682\\
Kapteyn's b & 0.008980084 & 0.00500124\\
Kepler-22 b & 0.01243731 & 0.007181929\\
Kepler-61 b & 0.011438662 & 0.006546287\\
Kepler-62 e	& 0.008842245 & 0.004915399\\
Kepler-62 f	& 0.007887122 & 0.004326487\\
Kepler-174 d & 0.011588827 & 0.006641471 \\
Kepler-186 f & 0.006442599 & 0.003450367 \\
Kepler-283 c & 	0.009799112 & 0.005511735\\
Kepler-296 f & 0.009704721 & 0.005452561 \\
Kepler-298 d & 0.013193284 & 0.007666263 \\
GJ 667C c & 0.007028218 & 0.00775173 \\
GJ 163 c & 0.022843579 & 0.005571684 \\
\hline
\end{tabular}
\end{table}

\begin{figure}
\centering
\begin{subfigure}[h!]{0.49\textwidth}
\centering
\includegraphics[width=1.1\textwidth,height=0.4\textheight]{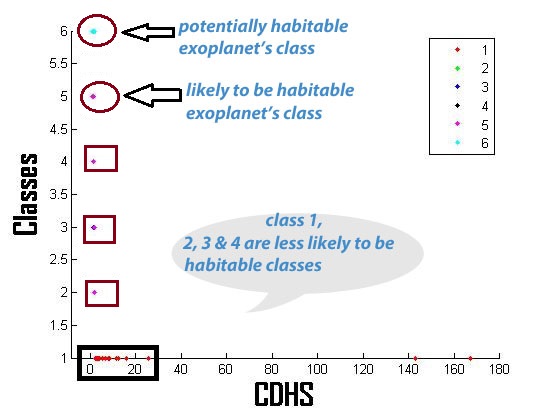}
\caption{for DRS on non-normalized data set}
\vskip 0.3in
\includegraphics[width=1.1\textwidth,height=0.4\textheight]{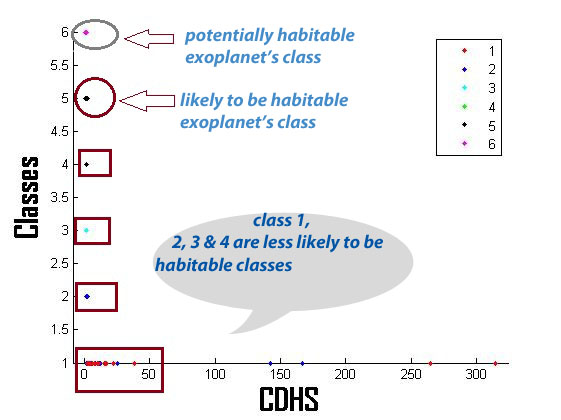}
\caption{for CRS on non-normalized data set}
\label{fig:KNNplot}
\end{subfigure}
\begin{subfigure}[h!]{0.49\textwidth}
\centering
\includegraphics[width=1.2\textwidth,height=0.4\textheight]{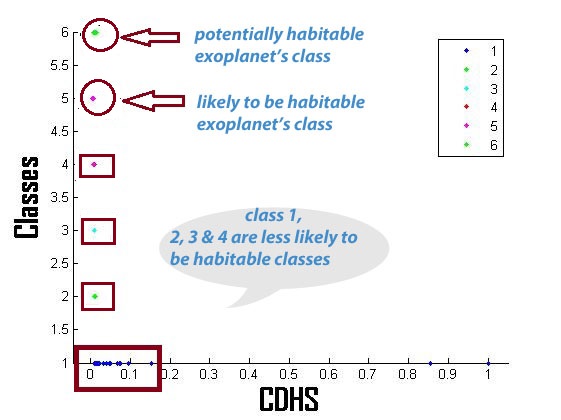}
\caption{for DRS on normalized data set}
\vskip 0.3in
\includegraphics[width=1.2\textwidth,height=0.4\textheight]{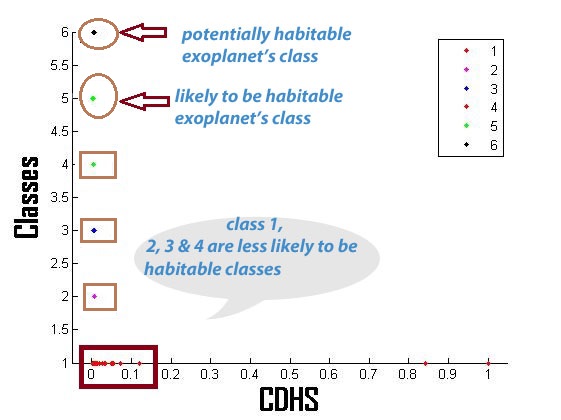}
\caption{for CRS on normalized data set}
\label{fig:KNNnormplot}
\end{subfigure}
\caption{Results of attribute enhanced K-NN algorithm. The $X$-axis represents the Cobb-Douglas habitability score and $Y$-axis -- the 6 classes: schematic representation of the outcome of our algorithm. The points in circles and boxes indicate membership in respective classes. These points are representative of membership only and do not indicate a quantitative equivalence of the exact representation. Full catalog is available at our website {\tt https://habitabilitypes.wordpress.com/}} 

\end{figure}

As observed, the results of classification are almost similar for non-normalized (Figs.~3a \& 3b) 
and normalized (Figs.~3c \& 3d) CDHS. Both methods have identified the exoplanets that were previously assumed as potentially habitable (listed in the HEC database) with comparable accuracy. However, after normalization, the accuracy increases from 94\% for $K=1$ to above 99\% for $K=7$. All our results for confirmed exoplanets from PHL-EC, including DRS and CRS habitability CDHS scores and classes assignations, are presented in the catalog at {\tt https://habitabilitypes.wordpress.com/}. CRS gave better results compared to DRS case in the non-normalized dataset, therefore, the final habitability score is considered to be the CDHS obtained in the CRS phase.\\  

\textbf{Remark:} Normalized and non-normalized CDHS are obtained by two different methods. After applying the K-NN on the non-normalized CDHS, the method produced $12$ and $14$ habitable exoplanets in CRS and DRS cases, respectively, from a list of $664$ exoplanets. The "Earth-League", Class $6$, is the class where the algorithm "dumps" those exoplanets which satisfy the conditions of K-NN and threshold and probabilistic herding as explained in Sections~$3.1$, $3.2$ and $3.3$. We applied this algorithm 
 again to the normalized CDHS of $664$ exoplanets under the same conditions. It is observed that the output was $16$ exoplanets that satisfied the conditions of being in Class 6, the "Earth-league", irrespective of CRS or DRS conditions. The reason is that the normalized scores are tighter and much closer to each other compared to the non-normalized CDHS, and that yielded a few more exoplanets in Class $6$.
 
{\bf ESI is a number that tells us whether an exoplanet is similar to Earth. It is common wisdom not to consider ESI as the only metric for habitability (\textbf{Rita, please explain the physical reasons behind}). Another metric, PHI can't be used as a single benchmark for habitability and a lot of other physical conditions have to be checked before a conclusion may be drawn.} The proposed scoring method outperforms both ESI and PHI in terms of accurately classifying exoplanets as habitable or not. The novel method of computing habitability by CD-HPF and CDHS, coupled with K-NN with probabilistic herding makes it feasible to use a single metric that assigns majority of the habitable exoplanets, in Earth's league. The existing K-NN algorithm has been modified and attribute enhanced voting scheme and probabilistic proximity have been utilized as a checkpoint for final class distributions -- we call it as the "Earth-League". For large enough data samples, there are theoretical guarantees that the algorithm will converge to a finite number of discriminating classes. The members of "Earth-League" are cross-validated with the list of potentially habitable exoplanets in the HEC database. The results (Table~9) render the proposed metric \textbf{CDHS} to behave with a reasonable degree of reliability.

\section{Conclusion and Future Work}

The two existing indices, ESI and PHI, are unreliable and sometimes controversial metrics as far as habitability of a planet is concerned. As observed, these values are not benchmarks for determination of habitability. At any rate, a benchmark of habitability may sound a bit ambitious given the perpetual complexity of the problem. This is the reason that other physical parameters such as planets being terrestrial or not, conditions to support existence of liquid water, have to be considered. This is where the CD-HPF model triumphs. The model generates 12 potentially habitable exoplanets in Class 6, which is considered to be a class where Earth-like planets reside. All these 12 exoplanets are identified as habitable by the PHL. The score generated by our model is a single metric which could be used to classify habitability of exoplanets as members of the "Earth League", unlike ESI and PHI. Attribute enhanced K-NN algorithm, implemented in the paper, helps achieve this goal and membership of exoplanets to different classes of habitability may change as the four input parameters of Cobb-Douglas model change values.

CD-HPF is a novel metric of defining habitability score for exoplanets. It needs to be noted that the authors perceive habitability as a probabilistic measure, or a measure with varying degrees of certainty. Therefore, the construction of different classes of habitability 1 to 6 is contemplated, corresponding to measures as ``most likely to be habitable" as Class 6, to ``least likely to be habitable" as Class 1. As a further illustration, classes 6 and 5 seem to represent the identical patterns in habitability, but they do not! Class 6 -- the "Earth-League", is different from Class 5 in the sense that it satisfies the additional conditions of thresholding and probabilistic herding and, therefore, ranks higher on the habitability score. This is in stark contrast to the binary definition of exoplanets being ``habitable or non-habitable", and a deterministic perception of the problem itself. The approach therefore required classification methods that are part of machine learning techniques and convex optimization --- a sub-domain, strongly coupled with machine learning. Cobb-Douglas function and CDHS are used to determine habitability and the  maximum habitability score of all exoplanets with confirmed surface temperatures in the PHL-EC. Global maxima is calculated theoretically and algorithmically for each exoplanet, exploiting intrinsic concavity of CD-HPF and ensuring "no curvature violation". Computed scores are fed to the attribute enhanced K-NN algorithm --- a novel classification method, used to classify the planets into different classes to determine how similar an exoplanet is to Earth. The authors would like to emphasize that, by using classical K-NN algorithm and not exploiting the probability of habitability criteria, the results obtained were pretty good, having 12 confirmed potentially habitable exoplanets in the "Earth-League". A catalog was created by the authors which lists the confirmed exoplanets with the class assignments and computed habitability scores. This catalog is built with the intention of further use in designing statistical experiments for the analysis of the correlation between habitability and the abundance of elements (this work is briefly outlined in Safonova et al., 2016). The full customized catalog of all confirmed exoplanets with class annotations is available at {\tt https://habitabilitypes.wordpress.com/}. It is a very important observation that our algorithm and methods give rise a score structure, CDHS, which is structurally similar to PHI as a corollary in the CRS case (when the elasticities in CRS are assumed to be equal to each other). Both are geometric means of the input parameters considered for the respective models.

CD-HPF uses four parameters (radius, density, escape velocity and surface temperature) to compute habitability score, which alone are not sufficient to determine habitability of exoplanets. Other parameters, such as e.g. orbital period, stellar flux, distance of the  planet from host star, etc. may be equally important to determine the habitability. Since our model is scalable, additional parameters can be added to achieve better and granular habitability score. In addition, out of $1900$ confirmed exoplanets in HEC, only 664 planets have their surface temperatures listed. For many expolanets, the surface temperature, which is an important parameter in this problem, is not known or not defined. The unknown surface temperatures can be estimated using various statistical models. Future work may include incorporating more input parameters, such as orbital velocity, orbital eccentricity, etc. to the Cobb-Douglas function, coupled with tweaking the attribute enhanced K-NN algorithm by checking an additional condition such as, e.g. distance to the host star. Cobb-Douglas, as proved, is a scalable model and doesn't violate curvature with additional predictor variables. However, it is pertinent to check for the dominant parameters that contribute more towards the habitability score. This can be accomplished by computing percentage contributions to the response variable -- the  habitability score. We would like to conclude by stressing on the efficacy of the method of using a few of the parameters rather than sweeping through a host of properties listed in the catalogs, effectively reducing the dimensionality of the problem. To sum up, CD-HPF and CDHS turn out to be self-contained metrics for habitability.\\

\textbf{Note: All relevant data and results: sets, figures, animation video and a graphical abstract, is available at our website, specially designed for this project, at {\tt  https://habitabilitypes.wordpress.com/}.
}

\vskip 0.1in 
\section*{Acknowledgments}

This research has made use of the PHL's Exoplanet Catalog maintained by the Planetary Habitability Laboratory at the University of Puerto Rico, Arecibo, and NASA Astrophysics Data System Abstract Service.

\newpage
\noindent
\appendix
\renewcommand{\thesection}{\Alph{section}}
\chapter{\textbf{Appendices}
}\section{\textbf{Special case: Heuristic for elasticity computation}
}
\numberwithin{equation}{section}
 
Let us consider the CD-HPF for gaining more insight to computing the elasticity for maximization of \textbf{CDHS} \citep{Snehanshu2016RevenueOpt}. The following heuristic produces easy and quick way to compute elasticity in real time.
\begin{equation}
Y=A^{\alpha}B^{\beta}\,,
\end{equation}
where $A$ and $B$ are constants. Let \([\alpha_{min},\alpha_{max}]\) be the range of permissible values for $\alpha$ and, similarly, $[\beta_{min},\beta_{max}]$ the range of permissible values for $\beta$, where \(\alpha_{min},\alpha_{max},\beta_{min},\beta_{max}>0\). To maximize \(Y\), if  \(A>1\) then \(\alpha=\alpha_{max}\) (\(\alpha\) should be as large as possible and \(\alpha_{max}\) is the largest permitted value). Similarly, if \(A<1\), then $\alpha=\alpha_{min}$.  Since the terms involving \(\alpha\) are independent of those involving $\beta$, the same logic can be applied independently to the term \(B^\beta\). An easy way to see the above is by taking log of both sides of (A.1), we get
\begin{equation}
\log{Y}=\alpha \log{A}+\beta \log{B}\,.
\end{equation}
To maximize $\log{Y}$, if $\log{A}$ is negative, \(\alpha\) needs to be as small as possible (since \(\alpha > 0\)) else \(\alpha\)
must be as large as possible. The same is applied to \(\beta\).

Consider the case where we have a set of data points, i.e. instead of constants $A$ and $B$, we have
\begin{equation}
y_i=u_i^{\alpha}v_i^{\beta} \,,
\end{equation}
where i=1 to N. 

Suppose our criterion is to choose \(\alpha\) and \(\beta\) so as to maximize $Y=\prod_{i=1}^{N}y_i$, i.e. maximize
\begin{equation}
\prod_{i=1}^{N}y_i=\left(\prod_{i=1}^{N}u_i\right)^\alpha\left(\prod_{i=1}^{N}v_i\right)^\beta\,.
\end{equation}
The RHS is similar in form to the essential CD function and, hence, same rule can be applied i.e. If \(\prod_{i=1}^{N}u_i<1\) then  \(\alpha=\alpha_{min}\) else \(\alpha=\alpha_{max}\).
The term involving \(\beta\) can be minimized similarly and independently. The only remaining step is to determine the permissible ranges. Let \(\epsilon\) be the smallest value
that \(\alpha\) and \(\beta\) can take.
Suppose in the above example, $\prod_{i=1}^{N} u_i<1$ and \(\prod_{i=1}^{N}v_i>1\). We know that \(\alpha\) should be minimized
and \(\beta\) should be maximized.
Since \(\alpha+\beta<1\), let \(\alpha+\beta=1-\delta\), where \(\delta\) is a small non-negative number. We then have \(\alpha_{min}=\epsilon\)
and \(\beta_{max}=1-\delta-\epsilon\). 

\section{Proof of optimization using Lagrangian multiplier }
\numberwithin{equation}{section}

The production maximization is done using Lagrangian multipliers. The Lagrangian function for the optimization problem is
\begin{align}
\cal L&= Y-\lambda(w_1R+w_2D+w_3T_{s}+w_4V_{e}-m)\,;\nonumber\\
\cal L&=kR^\alpha D^\beta T_{s}^\gamma V_{e}^\delta-\lambda(w_1R+w_2D+w_3T_{s}+w_4V_{e}-m)\,.\nonumber
\end{align}
The first order conditions are
\begin{align}&\frac{\partial \cal L}{\partial R}= k\alpha R^{\alpha-1}D^\beta T_{s}^\gamma V_{e}^\delta-w_1\lambda =0\\
&\frac{\partial \cal L}{\partial D}=k\beta R^{\alpha}D^{\beta-1} T_{s}^\gamma V_{e}^\delta-w_2\lambda  =0\\
&\frac{\partial \cal L}{\partial T_{s}}=k\gamma R^{\alpha}D^\beta T_{s}^{\gamma-1} V_{e}^\delta-w_3\lambda  =0\\
&\frac{\partial \cal L}{\partial V_{e}}=k\delta R^{\alpha}D^\beta T_{s}^\gamma V_{e}^{\delta-1}-w_4\lambda  =0\\
&\frac{\partial \cal L}{\partial \lambda}=-(w_1R+w_2D+w_3T_{s}+w_4V_{e}-m)=0\end{align}
Performing calculations the following values of R, D, \(T_{s}\) and \(V_{e}\) are obtained:\begin{align}
R=\left(pk\alpha^{1-\left(\beta+\gamma+\delta\right)}\beta^{\beta}\gamma^{\gamma}\delta^{\delta}w_1^{\beta+\gamma +\delta-1}w_2^{-\beta}w_3^{-\gamma}w_4^{-\delta}\right)^\frac{1}{1-\left(\alpha+\beta+\gamma+\delta\right)}\\
D=\left(pk\alpha^{\alpha}\beta^{1-\left(\alpha+\gamma+\delta\right)}\gamma^{\gamma}\delta^{\delta}w_1^{-\alpha}w_2^{\alpha+\gamma +\delta-1}w_3^{-\gamma}w_4^{-\delta}\right)^\frac{1}{1-\left(\alpha+\beta+\gamma+\delta\right)}\\
T_{s}=\left(pk\alpha^{\alpha}\beta^{\beta}\gamma^{1-\left(\alpha+\beta+\delta\right)}\delta^{\delta}w_1^{-\alpha}w_2^{-\beta}w_3^{\alpha+\beta +\delta-1}w_4^{-\delta}\right)^\frac{1}{1-\left(\alpha+\beta+\gamma+\delta\right)}\\
V_{e}=\left(pk\alpha^{\alpha}\beta^{\beta}\gamma^{\gamma}\delta^{1-\left(\alpha+\beta+\gamma\right)}w_1^{-\alpha}w_2^{-\beta}w_3^{-\gamma}w_4^{\alpha+\beta +\gamma-1}\right)^\frac{1}{1-\left(\alpha+\beta+\gamma+\delta\right)}
\end{align}
Dividing (B.7), (B.8), (B.9) by (B.6), the following simplified expressions are obtained:
\begin{align*}
D&=\frac{\beta}{\alpha}\frac{w_1}{w_2}R\\
T_{s}&=\frac{\gamma}{\alpha}\frac{w_1}{w_3}R\\
V_{e}&=\frac{\delta}{\alpha}\frac{w_1}{w_4}R
\end{align*}
These expressions will be observed in the subsequent part of the proof again!
The Lagrangian function for the optimization problem is:
\begin{align}
\cal L&= w_1R+w_2D+w_3T_{s}+w_4V_{e}-\lambda(f(R,D,T_{s},V_{e})-y_{tar})\,.
\end{align}
The first-order conditions are;
\begin{align}&\frac{\partial \cal L}{\partial R}=w_1-\lambda  k\alpha R^{\alpha-1} D^\beta T_{s}^\gamma V_{e}^\delta=0\,
\end{align}
\begin{align}&\frac{\partial \cal L}{\partial D}=w_2-\lambda  k\beta R^\alpha D^{\beta-1} T_{s}^\gamma V_{e}^\delta=0\,
\end{align}
\begin{align}&\frac{\partial \cal L}{\partial T_{s}}=w_3-\lambda  k\gamma R^\alpha D^\beta T_{s}^{\gamma-1} V_{e}^\delta=0\,
\end{align}
\begin{align}&\frac{\partial \cal L}{\partial V_{e}}=w_4-\lambda  k\delta R^\alpha D^\beta T_{s}^\gamma V_{e}^{\delta-1}=0\,
\end{align}
\begin{align}&\frac{\partial \cal L}{\partial \lambda}=k R^\alpha D^\beta T_{s}^\gamma V_{e}^\delta-y_{tar}=0\,.
\end{align}

Substituting the values of the above 4 parameters in equation (B.10), we get
\begin{align}
\Rightarrow& \, y_{tar}=kR^{\alpha}\left(\frac{\beta}{\alpha}\frac{w_1}{w_2}R\right)^\beta \left(\frac{\gamma}{\alpha}\frac{w_1}{w_3}R\right)^\gamma\left(\frac{\delta}{\alpha}\frac{w_1}{w_4}R\right)^\delta\nonumber \\
\Rightarrow& \, y_{tar}=kR^{\alpha+\beta+\gamma+\delta}\alpha^{-\beta-\gamma-\delta}\beta^{\beta}\gamma^{\gamma}\delta^{\delta}w_1^{\beta+\gamma+\delta}w_2^{-\beta}w_3^{-\gamma}w_4^{-\delta}\nonumber\\
\Rightarrow& \, R^{\alpha+\beta+\gamma+\delta}=k^{-1}\alpha^{\beta+\gamma+\delta}\beta^{-\beta}\gamma^{-\gamma}\delta^{-\delta}w_1^{-\beta-\gamma-\delta}w_2^{\beta}w_3^{\gamma}w_4^{\delta}y_{tar}\nonumber\\
\Rightarrow& \, R=\left(k^{-1}\alpha^{\beta+\gamma+\delta}\beta^{-\beta}\gamma^{-\gamma}\delta^{-\delta}w_1^{-\beta-\gamma-\delta}w_2^{\beta}w_3^{\gamma}w_4^{\delta}y_{tar}\right)^\frac{1}{\alpha+\beta+\gamma+\delta}\nonumber\\
\Rightarrow& \, w_1R=\left(k^{-1}\alpha^{\beta+\gamma+\delta}\beta^{-\beta}\gamma^{-\gamma}\delta^{-\delta}w_1^{\alpha}w_2^{\beta}w_3^{\gamma}w_4^{\delta}y_{tar}\right)^\frac{1}{\alpha+\beta+\gamma+\delta}\,.
\end{align}
Similarly,
\begin{align}
w_2D=\left(k^{-1}\alpha^{-\alpha}\beta^{\beta+\gamma+\delta}\gamma^{-\gamma}\delta^{-\delta}w_1^{\alpha}w_2^{\beta}w_3^{\gamma}w_4^{\delta}y_{tar}\right)^\frac{1}{\alpha+\beta+\gamma+\delta}\\
w_3T_{s}=\left(k^{-1}\alpha^{-\alpha}\beta^{-\beta}\gamma^{\beta+\gamma+\delta}\delta^{-\delta}w_1^{\alpha}w_2^{\beta}w_3^{\gamma}w_4^{\delta}y_{tar}\right)^\frac{1}{\alpha+\beta+\gamma+\delta}\\
w_4V_{e}=\left(k^{-1}\alpha^{-\alpha}\beta^{-\beta}\gamma^{-\gamma}\delta^{\beta+\gamma+\delta}w_1^{\alpha}w_2^{\beta}w_3^{\gamma}w_4^{\delta}y_{tar}\right)^\frac{1}{\alpha+\beta+\gamma+\delta}
\end{align}
The cost for producing \(y_{tar}\) units in cheapest way is c, where
\begin{align}
c=w_1R+w_2D+w_3T_{s}+w_4V_{e}\,.
\end{align}
Analytical representation of $c$ can be rewritten from Eq.~(B.20) as
\begin{align}
c=Q\left[w_1^\alpha w_2^\beta w_3^\gamma w_4^\delta \right]^\frac{1}{\alpha+\beta+\gamma+\delta}y_{tar}^\frac{1}{\alpha+\beta+\gamma+\delta}
\,,\end{align}
where 
$$
Q=k^\frac{-1}{\alpha+\beta+\gamma+\delta}\left[\frac{\alpha^{\beta+\gamma+\delta}}{\beta^\beta+\gamma^\gamma+\delta^\delta}+\frac{\beta^{\alpha+\gamma+\delta}}{\alpha^\alpha+\gamma^\gamma+\delta^\delta}+\frac{\gamma^{\alpha+\beta+\delta}}{\alpha^\alpha+\beta^\beta+\delta^\delta}+\frac{\delta^{\alpha+\beta+\gamma}}{\alpha^\alpha+\beta^\beta+\gamma^\gamma}\right]^\frac{1}{\alpha+\beta+\gamma+\delta}\,,
$$
with
$$
c_{avg}=\frac{c}{y_{tar}}=Q\left[w_1^\alpha w_2^\beta w_3^\gamma w_4^\delta \right]^\frac{1}{\alpha+\beta+\gamma+\delta}y_{tar}^{\frac{1}{\alpha+\beta+\gamma+\delta}-1}\,.
$$

Deriving the conditions for optimization:
\begin{align}
p\alpha kR^{\alpha-1}D^\beta T_{s}^\gamma V_{e}^\delta =w_1\\
p\beta kR^{\alpha}D^{\beta-1} T_{s}^\gamma V_{e}^\delta =w_2\\
p\gamma kR^{\alpha}D^\beta T_{s}^{\gamma-1} V_{e}^\delta =w_3\\
p\delta kR^{\alpha}D^\beta T_{s}^\gamma V_{e}^{\delta-1} =w_4\end{align}
Multiplying these equations with R, D,\(T_{s}\) and \(V_{e}\), respectively,

\begin{align}
p\alpha kR^{\alpha}D^\beta T_{s}^\gamma V_{e}^\delta =w_1R           &\Rightarrow   p\alpha Y=w_1R\\
p\beta kR^{\alpha}D^{\beta} T_{s}^\gamma V_{e}^\delta =w_2D   &\Rightarrow    p\beta Y=w_2D\\
p\gamma kR^{\alpha}D^\beta T_{s}^{\gamma} V_{e}^\delta =w_3T_{s}   &\Rightarrow    p\gamma Y=w_3T_{s}\\
p\delta kR^{\alpha}D^\beta T_{s}^\gamma V_{e}^{\delta} =w_4V_{e}   &\Rightarrow    p\delta Y=w_4V_{e}
\end{align}
Dividing equations (B.27), (B.28) and (B.29) by (B.26) following equations are obtained:
\begin{align}
D&=\frac{\beta}{\alpha}\frac{w_1}{w_2}R\\
T_{s}&=\frac{\gamma}{\alpha}\frac{w_1}{w_3}R\\
V_{e}&=\frac{\delta}{\alpha}\frac{w_1}{w_4}R
\end{align}
Substituting these values of $D$, \(T_{s}\) and \(V_{e}\) into Eq.~(B.26) and performing some simple algebraic calculations, we obtain
\begin{align}
&p\alpha kR^{\alpha-1}D^\beta T_{s}^\gamma V_{e}^\delta=w_1\nonumber \\
\Rightarrow & \,p\alpha kR^{\alpha-1}\left(\frac{\beta}{\alpha}\frac{w_1}{w_2}R\right)^\beta \left(\frac{\gamma}{\alpha}\frac{w_1}{w_3}R\right)^\gamma\left(\frac{\delta}{\alpha}\frac{w_1}{w_4}R\right)^\delta=w_1\nonumber \\
\Rightarrow &\, pkR^{\alpha+\beta+\gamma+\delta -1}\beta^{\beta}\gamma^{\gamma}\delta^{\delta}w_1^{\beta+\gamma+\delta-1}w_2^{-\beta}w_3^{-\gamma }w_4^{-\delta}=1\nonumber \\
\Rightarrow&\, R=\left(pk\alpha^{1-\left(\beta+\gamma+\delta\right)}\beta^{\beta}\gamma^{\gamma}\delta^{\delta}w_1^{\beta+\gamma +\delta-1}w_2^{-\beta}w_3^{-\gamma}w_4^{-\delta}\right)^\frac{1}{1-\left(\alpha+\beta+\gamma+\delta\right)}\,.
\end{align}
After performing similar calculations, the following expressions of $D$, \(T_{s}\) and \(V_{e}\) are obtained:
\begin{align}
D&=\left(pk\alpha^{\alpha}\beta^{1-\left(\alpha+\gamma+\delta\right)}\gamma^{\gamma}\delta^{\delta}w_1^{-\alpha}w_2^{\alpha+\gamma +\delta-1}w_3^{-\gamma}w_4^{-\delta}\right)^\frac{1}{1-\left(\alpha+\beta+\gamma+\delta\right)}\\
T_{s}&=\left(pk\alpha^{\alpha}\beta^{\beta}\gamma^{1-\left(\alpha+\beta+\delta\right)}\delta^{\delta}w_1^{-\alpha}w_2^{-\beta}w_3^{\alpha+\beta +\delta-1}w_4^{-\delta}\right)^\frac{1}{1-\left(\alpha+\beta+\gamma+\delta\right)} \\
V_{e}&=\left(pk\alpha^{\alpha}\beta^{\beta}\gamma^{\gamma}\delta^{1-\left(\alpha+\beta+\gamma\right)}w_1^{-\alpha}w_2^{-\beta}w_3^{-\gamma}w_4^{\alpha+\beta +\gamma-1}\right)^\frac{1}{1-\left(\alpha+\beta+\gamma+\delta\right)} 
\end{align}
These values of $R$, $D$, $T_{s}$ and $V_{e}$ are the expressions to be maximized. Substituting values of $R$, $D$, \(T_{s}\) and \(V_{e}\) into CD-HPF, \begin{equation}
Y=f\left(R,D,T_{s},V_{e}\right)=\left(R\right)^{\alpha}\cdot \left(D\right)^{\beta}
\cdot \left(T_{s}\right)^{\gamma}\cdot\left(V_{e}\right)^{\delta}\,,
\end{equation} 
we obtain
\begin{align}
Y=\left(kp^{\alpha+\beta+\gamma+\delta}\alpha^{\alpha}\beta^{\beta}\gamma^{\gamma}\delta^{\delta}w_1^{-\alpha}w_2^{-\beta}w_3^{-\gamma}w_4^{-\delta}\right)^\frac{1}{1-\left(\alpha+\beta+\gamma+\delta\right)}\,.
\end{align}
If $\alpha+\beta+\gamma+\delta < 1$, the exponent on the right hand side of the above equation remains strictly positive and $Y$, the habitability score, increases in a bounded fashion. This is a natural extension to the sample 3D CD-HPF model for DRS, where the constraint in two input parameters is $\alpha+\beta<1$ (please refer to Matlab codes in Appendix~D).

\section{Hessian Matrix: Conditions for concavity for CRS and DRS}

\numberwithin{equation}{section}
A \(c^2\) function \(f:U\subset R^n\rightarrow R\)  defined on a convex open set $U$ is concave if and only if the Hessian matrix \(D^2f(x)\) is negative semi-definite for all \(x\in U\). A matrix $H$ is negative semi-definite if and only if its \(2^n-1\) principal minors alternate in sign, so that odd order minors are less than equal to 0, and even order minors are greater than equal to 0. The  Cobb-Douglas function for two inputs is:
$$
Y==f(x,y)=kx_1^{\alpha}x_2^{\beta}\,.
$$
Its Hessian is 
$$
\begin{bmatrix}\alpha(\alpha-1)kx_1^{\alpha-2}x_2^\beta&\alpha\beta kx_1^{\alpha -1}x_2^{\beta-1}\\\alpha\beta kx_1^{\alpha -1}x_2^{\beta-1}&\beta(\beta-1)kx_1^{\alpha}x_2^{\beta-2}
\end{bmatrix}\,,
$$
where
\begin{align}
\Delta_1&=\alpha(\alpha-1)kx_1^{\alpha-2}x_2^\beta\nonumber\\ 
\Delta_1&=\beta(\beta-1)kx_1^{\alpha}x_2^{\beta-2}\nonumber\\ 
\Delta_2&=\alpha\beta k^2x_1^{2\alpha-2}x_2^{2\beta-2}(1-(\alpha+\beta))\,.\nonumber
\end{align}
For DRS and CRS, \(\alpha  + \beta \leq 1\) and $\alpha \geq 0$, $\beta \geq 0$\nonumber. Since all other terms in $\Delta_2 $ are greater than $0$, and  
\begin{align}
&(1-(\alpha+\beta))\geq0\nonumber\\  
\Rightarrow & \, \Delta_2\geq 0\,.\nonumber
\end{align}
By inspection, $\alpha(\alpha-1)$ and $\beta(\beta-1)$ are less than or equal to $0$. Other terms in $\Delta_1$ are non-negative and hence the product, 
$$
\Delta_1\leq 0\,.
$$
Thus, conditions for CD-HPF to be concave, i.e. 
\begin{align*}
&\Delta_1\leq 0\\ 
&\Delta_2\geq 0\,.
\end{align*} 
are satisfied by DRS and CRS.
This is in agreement with the graphs obtained for DRS and CRS; while for IRS the graph is neither concave nor convex. Therefore, no formulation of CD-HPF and subsequent computation for CDHS involves the IRS phase.

\section{MATLAB Codes}

\section*{Function fmincon}

\noindent
The function {\em fmincon} finds a constrained minimum of a scalar function of multivariable starting at an initial point. This is generally known as constrained nonlinear optimization. Function {\em fmincon} solves problems of the form: \\
min $f(x)$ subject to $x$,
\begin{equation}
\begin{cases}
A*x \leq b   \\
A_{eq} *x=b_{eq}
\end{cases}\nonumber
\end{equation}
are the linear constraints, and the following equations are the non-linear constraints: 
\begin{equation}
\begin{cases}
C*x\leq 0\\
C_{eq} *x=0
\end{cases}\nonumber
\end{equation}\\
and bounding of variables	
\begin{equation}
\begin{cases}
lb  \leq  x\\ 
x  \leq  ub
\end{cases}\nonumber
\end{equation}
This has been applied to the cases, \textbf{CRS} and \textbf{DRS} for the CD-HPF and CDHS computation. The trick to using \emph{fmincon} lies in computing the elasticities $\alpha$ and $\beta$ of CRS and DRS in the context of a sample 3D CD-HPF. The values of elasticities, thus obtained, help optimize CDHS for each exoplanet.

\subsection*{\textbf{Constant Returns to Scale}}
\noindent 
Applying the constraints: 
\begin{equation}
\begin{cases}
\alpha  + \beta =1\\
\alpha>0,\, \beta>0 
\end{cases}\nonumber
\end{equation}
to the function:
\(Y=kx_1^{\alpha}x_2^{\beta}\nonumber\); use \emph{fmincon} to compute $\alpha$  and $\beta$ for optimum $Y$.

\subsection*{\textbf{Decreasing Returns to Scale}}
\noindent
Applying the constraints:
\begin{equation}
\begin{cases}
\alpha  + \beta < 1\\
\alpha>0, \,\beta>0 
\end{cases}\nonumber
\end{equation}
to the function: \(Y=kx_1^{\alpha}x_2^{\beta}\nonumber\); use \emph{fmincon} to compute $\alpha$ and $\beta$ for optimum $Y$.\\  

\noindent
\textbf{NOTE:} Identical technique is employed to compute elasticity values, $\delta$ and $\gamma$ for the scaled up model, 
$$
Y=kx_1^{\alpha}x_2^{\beta}x_3^{\delta}x_4^{\gamma}\,.
$$

\section*{Syntax of fmincon}

\noindent
[x,fval] = fmincon(fun,$x_{0},A,b$) starts at point $x_{0}$ and finds a minimum $x$ to the function described in fun subject to the linear inequalities, $A*x \leq b$, where $A$ is a matrix, $x$ and $b$ are vectors and $x_{0}$ can be a scalar, a vector or a matrix. It also returns the value of the objective function \textbf{fun} at the solution $x$.
\newline
\newline
[x,fval] = fmincon(fun,$x_{0},A,b,A_{eq},b_{eq}$) starts at $x_{0}$ and minimizes \textbf{fun} subject to the linear inequalities $A_{eq}*x = b_{eq}$ and $A*x \leq b$, where $A_{eq}$ is a matrix and $b_{eq}$ is a vector. It also returns the value of the objective function \textbf{fun} at the solution $x$. \\
\newline
Function {\em fmincon()} has four algorithm options: 
\begin{itemize}
	\item{interior-point} 
	\item{sqp}	
    \item{active-set}
	\item{trust-region-reflective}
\end{itemize}
Trust-region-reflective is the default algorithm uses by {\em fmincon}. In our case, we have also used the default one.

\section*{Matlab code for Decreasing Returns to Scale:}

\noindent
\(x_{0}\) = [0.2,0.2];\newline
A = [1 1;-1 0;0 -1];\newline
b = [0.9;-0.1;-0.1];\newline
[x,fval] = fmincon(@cobb,\(x_{0},A,b\));\newline
function f = cobb(x); \newline
/*where f is the outcome of Cobb-Douglas function and $x(1)$ and $x(2)$ are the elasticities*/
\begin{equation}
f = -1.99 ^{x(1)} .* 1.06 ^{ x(2)}; \nonumber 
\end{equation}
end

\section*{3D plot code for DRS}

\noindent
syms xm ym;\newline
$N=663$;\newline
$dy=0.001$;\newline
$dx=0.001$;\newline
[xm,ym] = meshgrid($0.1:dx:0.9,0.1:dy:0.9$);\newline
\begin{equation}
f=-1.57 ^ {xm} .*  573.18 ^{ym};\nonumber \end{equation} 
$f(xm+ym>0.9) =$ NaN;\newline
surf(xm,ym,f,'EdgeColor','none'); 

\section*{Matlab code for Constant Returns to Scale:}

\noindent
$x_{0} = [0.4,0.2]$;\newline
$A = [-1 0;0 -1]$;\newline
$b = [-0.1;-0.1]$;\newline
$A_{eq} = [1 1]$;\newline
$b_{eq} = [1]$;\newline
$[x,fval] =$ fmincon(@cobb,$x_{0},A,b,A_{eq},b_{eq}$);\newline
\newline
function f = cobb(x); \newline
/*where f is the outcome of Cobb-Douglas function and x(1) and x(2) are the elasticities*/
\begin{equation}
f = -1.99 ^{x(1)} .* 1.06 ^{x(2)}; \nonumber 
\end{equation}
end

\section*{3D plot code for CRS}

\noindent
syms xm ym;\newline
$N=663$;\newline
$dy=0.001$;\newline
$dx=0.001$;\newline
[xm,ym] = meshgrid($0.1:dx:0.9,0.1:dy:0.9$);\newline
\begin{equation}f=-1.57 ^ {xm} .* 573.18 ^{ym};\nonumber \end{equation} 
$f(xm+ym>1)= $NaN;\newline
surf(xm,ym,f,'EdgeColor','none');

\section{Attribute-Enhanced K-NN Algorithm: pseudo code} 
Consider $K$ as the desired number of nearest neighbors and $S:=p_{1},\dots,p_{n}$ be the set of training samples in the form $p_{i}=(x_{i},c_{i})$, where \(x_{i}\) is the $d$-dimensional feature vector of the point \(p_{i}\) and \(c_{i}\) is the class that \(p_{i}\) belongs to. In our case, the dimension $d=1$. Similarly, set $S':=p_{1'},\dots,p_{m'}$ to be the set of testing samples. 

\noindent
 \begin{algorithmic}
 \STATE $N \leftarrow 664$ 	
   \STATE $M \leftarrow 530$
   \STATE $n \leftarrow 134$
   \STATE $boundary \leftarrow 1$\\
$threshold \leftarrow 1$\\
\end{algorithmic}
{\bf for $i=1$ to $N$ do\\
 if $habitability_{i}=1$\\
 \noindent 
 prob(\(habitability_{i}\)) = `high'\\
 else\\
 \noindent 
 prob(\(habitability_{i}\)) = `low'\\
for $i=1$ to $M$ do,\\
if (prob$(habitability_{i}) =$ `high' and \\ CDHS($p_{i}$)-CDHS(earth)$<=$ boundary)\\
\noindent
\(exoplanet_{i}\) belongs to Class 6\\
else\\
if CDHS of \(exoplanet_{i}\) falls in certain range\\
\noindent 
classify it accordingly in one of the remaining 5 classes  \\
for each $p'=(x',c')$ \\
 		Compute the distance $d(x',x_{i})$ between $p'$ and all $p_{i}$ belonging to $S$ \\
 		Select the $k$ nearest points to $p'$ from the list computed above \\
 		\emph{Apply Probabilistic Herding}: Assign a class to $p'$ based on the conditions
\begin{itemize}
\item if prob(\(habitability_{i}\)) = `high' and satisfies the boundary condition mentioned above
         assign class $c'$ to $p'$
\item else assign $p'$ the class according to the range set for each class.
\end{itemize}
}
\newpage

\section{Number of exoplanets in each class}

This section gives the statistics of the number of exoplanets in six classes for all the four cases. The tables below show the details of the class number with the number of exoplanets belong to each class.

\begin{table}[h!]
\begin{center}
\parbox{.45\linewidth}{
\begin{center}
\caption{Number of exoplanets in each class on DRS}
\begin{tabular}{c c } 
\hline 
Class Number&Number of exoplanets\\
[0.5ex]
\hline
6 & 14 \\
5 & 131\\
4 & 129\\
3 & 123\\
2 & 133 \\
1 & 133 \\
\hline
\end{tabular}
\end{center}
}
\hfill
\parbox{.45\linewidth}{
\begin{center}
\caption{Number of exoplanets in each class on CRS}
\begin{tabular}{c c} 
\hline 
 Class Number&Number of exoplanet \\
[0.5ex]
\hline
6 & 12 \\
5 & 138 \\
4 & 129 \\
3 & 126 \\
2 & 129\\
1 & 128 \\
\hline
\end{tabular}
\end{center}
}
\end{center}
\end{table}
\begin{table}[h!]
\begin{center}
\parbox{.45\linewidth}{
\begin{center}
\caption{Number of exoplanets in each class on DRS with normalized data}
\begin{tabular}{c c } 
\hline 
Class Number&Number of exoplanets\\
[0.5ex]
\hline
6 & 16 \\
5 & 130\\
4 & 129\\
3 & 125\\
2 & 129 \\
1 & 134 \\
\hline
\end{tabular}
\end{center}
}
\hfill
\parbox{.45\linewidth}{
\begin{center}
\caption{Number of exoplanets in each class on CRS with normalized data}
\begin{tabular}{c c} 
\hline 
 Class Number&Number of exoplanets \\
[0.5ex]
\hline
6 & 16 \\
5 & 129 \\
4 & 129 \\
3 & 126 \\
2 & 131\\
1 & 132 \\
\hline
\end{tabular}
\end{center}
}
\end{center}
\end{table}

\end{document}